\documentclass{elsart5p}
\usepackage{graphics}
\usepackage{graphicx}
\usepackage{epsfig}
\usepackage{amssymb}
\usepackage{amsmath}

\newcommand{\Szero}{\mbox{$^{1\!}S_0$}}
\renewcommand{\pol}[1]{\mathaccent"017E{#1}}
\def\fmn#1#2{\mbox{${\textstyle \frac{#1}{#2}}$}}
\begin{document}
\begin{frontmatter}

\title{Analysing powers and spin correlations in deuteron-proton charge exchange at 726~MeV}

\author[ikp,jinr1]{S.~Dymov\corauthref{cor1}}, \ead{s.dymov@fz-juelich.de}\corauth[cor1]{Corresponding author.}
\author[jinr1]{T.~Azaryan},
\author[ikp,hepi]{Z.~Bagdasarian},
\author[pnpi]{S.~Barsov},
\author[ipnup]{J.~Carbonell},
\author[hepi,ikp]{D.~Chiladze},
\author[ikp]{R.~Engels},
\author[ikp]{R.~Gebel},
\author[ikp,pnpi]{K.~Grigoryev},
\author[ikp]{M.~Hartmann},
\author[ikp]{A.~Kacharava},
\author[munster]{A.~Khoukaz},
\author[jinr1]{V.~Komarov},
\author[krakow]{P.~Kulessa},
\author[jinr1]{A.~Kulikov},
\author[jinr1]{V.~Kurbatov},
\author[hepi]{N.~Lomidze},
\author[ikp]{B.~Lorentz},
\author[hepi,jinr1]{G.~Macharashvili},
\author[hepi,ikp]{D.~Mchedlishvili},
\author[ikp,jinr1]{S.~Merzliakov},
\author[munster]{M.~Mielke},
\author[ikp,pnpi]{M.~Mikirtychyants},
\author[ikp,pnpi]{S.~Mikirtychyants},
\author[hepi]{M.~Nioradze},
\author[ikp]{H.~Ohm},
\author[ikp]{D.~Prasuhn},
\author[ikp]{F.~Rathmann},
\author[ikp]{V.~Serdyuk},
\author[ikp]{H.~Seyfarth},
\author[jinr1]{V.~Shmakova},
\author[ikp]{H.~Str\"oher},
\author[hepi]{M.~Tabidze},
\author[ikpros,skob]{S.~Trusov},
\author[jinr1]{D.~Tsirkov},
\author[jinr1,msu]{Yu.~Uzikov},
\author[pnpi,bonn]{Yu.~Valdau},
\author[ferr]{C.~Weidemann},
\author[ucl]{C.~Wilkin}

\address[ikp]{Institut f\"ur Kernphysik and J\"ulich Centre for Hadron Physics, Forschungszentrum J\"ulich, D-52425 J\"ulich, Germany}
\address[jinr1]{Laboratory of Nuclear Problems, JINR, RU-141980 Dubna, Russia}
\address[hepi]{High Energy Physics Institute, Tbilisi State University, GE-0186 Tbilisi, Georgia}
\address[pnpi]{High Energy Physics Department, Petersburg Nuclear Physics Institute, RU-188350 Gatchina, Russia}
\address[ipnup]{Institut de Physique Nucl\'{e}aire, Universit\'e Paris-Sud, IN2P3-CNRS, F-91406 Orsay Cedex, France}
\address[munster]{Institut f\"ur Kernphysik, Universit\"at M\"unster, D-48149 M\"unster, Germany}
\address[krakow]{H.~Niewodnicza\'{n}ski Institute of Nuclear Physics PAN, PL-31342 Krak\'{o}w, Poland}
\address[ikpros]{Institut f\"ur Kern- und Hadronenphysik, Forschungszentrum Rossendorf, D-01314 Dresden, Germany}
\address[skob]{Skobeltsyn Institute of Nuclear Physics, Lomonosov Moscow State University, RU-119991 Moscow, Russia}
\address[msu]{Department of Physics, M.~V.~Lomonosov Moscow State University, RU-119991 Moscow, Russia}
\address[bonn]{Helmholtz-Institut f\"ur Strahlen- und Kernphysik, Universit\"at Bonn, D-53115 Bonn, Germany}
\address[ferr]{University of Ferrara and INFN, I-44100 Ferrara, Italy}
\address[ucl]{Physics and Astronomy Department, UCL, Gower Street, London, WC1E 6BT, UK}
\date{\today}

\begin{abstract}
The charge exchange of vector polarised deuterons on a polarised hydrogen
target has been studied in a high statistics experiment at the COSY-ANKE
facility at a deuteron beam energy of $T_d=726$~MeV. By selecting two fast
protons at low relative energy $E_{pp}$, the measured analysing powers and
spin correlations are sensitive to interference terms between specific
neutron-proton charge-exchange amplitudes at a neutron kinetic energy of
$T_n\approx \frac{1}{2}T_d =363$~MeV. An impulse approximation calculation,
which takes into account corrections due to the angular distribution in the
diproton, describes reasonably the dependence of the data on both $E_{pp}$
and the momentum transfer. This lends broad support to the current
neutron-proton partial-wave solution that was used in the estimation.
\end{abstract}

\end{frontmatter}


It is a consequence of the nucleon spins that, assuming charge independence,
five complex amplitudes are needed to describe neutron-proton elastic
scattering~\cite{MCG1960}. This means that, above the pion production
threshold, at least nine independent measurements are required at each
scattering angle to allow an unambiguous partial wave decomposition. Some of
the resulting observables, which could depend on up to three spin
projections~\cite{BYS1978}, are difficult to determine and values may only be
obtained indirectly through combinations of other measurements.

It was shown several years ago~\cite{BUG1987} that, at small momentum
transfers between the deuteron and the diproton, the tensor analysing power
in the deuteron charge exchange on hydrogen, $\pol{d}p\to \{pp\}_{\!s}n$, is
closely linked to the spin transfer in neutron-proton large angle scattering,
$\pol{p}n\to\pol{n}p$, provided that the excitation energy $E_{pp}$ in the
final diproton is very low. Due to the Pauli principle the two protons are
then dominantly in the \Szero\ state with antiparallel spins, here denoted by
$\{pp\}_{\!s}$, so that there is then a spin-isospin flip to this state from
the initial deuteron.

Further information on the neutron-proton scattering amplitudes can be
obtained through measurements of the analysing powers and spin correlations
in the $\pol{d}\pol{p}\to \{pp\}_{\!s}n$ reaction and measurements of this
type were carried out at deuteron beam energies of $T_d=1.2$~GeV and 2.27~GeV
to investigate the $np$ amplitudes at neutron kinetic energies of $T_n\approx
\frac{1}{2}T_d = 600$ and 1135~MeV~\cite{MCH2013}. Both transverse spin
correlations and the proton and deuteron analysing powers were investigated
and the results were found to be consistent with modern partial wave
solutions~\cite{ARN2000} at $T_n=600$~MeV, while failing badly at 1135~MeV.
We report here on a similar investigation carried out at $T_d=726$~MeV in a
high statistics experiment, where tighter cuts could be placed on $E_{pp}$
and small effects could be studied in detail. Since one might expect that the
$np$ partial wave amplitudes should be fairly reliable at 363~MeV, this is
the ideal testing ground to establish quantitatively the validity of the
theoretical modelling of deuteron charge exchange~\cite{CAR1991}.

The experiment was undertaken using the ANKE magnetic spectrometer installed
at an internal target position of the Cooler Synchrotron (COSY) at the
Forschungszentrum J\"ulich~\cite{BAR1997}. Data were taken in parallel with
those used to determine the spin correlations in quasi-free
$\pol{n}\pol{p}\to \{pp\}_{\!s}\pi^-$~\cite{DYM2013} and more details of the
experimental procedure, in particular of the measurements of the beam and
target polarisations, are to be found in this reference.

Only deuteron beams with  vector polarisation were used in this experiment
and these had ideal values of $p_d^{\uparrow}=\frac{2}{3}$ and
$p_d^{\downarrow}=-\frac{2}{3}$. The polarisations measured at the injection
energy of 75.6~MeV with the low energy polarimeter were
$p_d^{\uparrow}=+61\pm 4$\% and $p_d^{\downarrow}=-50\pm 3$\% for the two
states, while the tensor polarisations were shown to be below 2\%.

In order to increase the luminosity in the experiment, a jet of polarised
atomic hydrogen was fed into a $25~\mu$ thick teflon-coated aluminum storage
cell target with dimensions $x\times y\times z = 15\times 19\times
390$~mm$^3$. Here the $y$-direction is perpendicular to the COSY plane and
the $x$-direction is in this plane but perpendicular to the beam ($z$)
direction. The polarisation, $p_p$, of the target was in the $y$ direction and
its sign was reversed every five seconds. The mean value of the polarisation
was determined through the study of the quasi-free $np\to d\pi^0$ asymmetry
to be $|p_p|=69\%\pm 2\% (\text{stat}) \pm 3.5\%\ (\text{syst})$. A more
precise value of the product of the magnitudes of the beam and target
polarisation was, however, extracted from an analysis of the pion production
data themselves, which gave an average of $|p_d||p_p|=0.373\pm
0.015$~\cite{DYM2013}\footnote{The dilution of the polarisation between the
deuteron and the constituent neutron was minimised by preferentially
selecting low Fermi momenta in the deuteron, as shown for the analogous case
in Fig.~2a of Ref.~\cite{DYM2013}.}.

Although the ANKE spectrometer is equipped with other elements, the only
detector used in the charge-exchange experiment was the forward detector (FD)
that identified and measured the two fast final protons from the
$\pol{d}\pol{p}\to \{pp\}n$ reaction or, for polarisation studies, the fast
deuteron and spectator proton from the $\pol{d}\pol{p}\to dp\pi^0$ reaction.
The FD comprises a set of multiwire proportional and drift chambers and a
two-plane scintillation hodoscope~\cite{DYM2004}.

Having registered two charged particles in the FD, the isolation of the
$dp\to ppn$ reaction depends on identifying these as protons on the basis of
time-of-flight criteria that are described in detail in Ref.~\cite{MCH2013}.
For this purpose the difference in the times of flight of the particles
recorded in the FD is compared to that calculated on the assumption that the
two particles are both protons. This procedure suppresses enormously the
background that is associated, for example, with deuteron-proton pairs coming
from $dp$ elastic scattering.

\begin{figure}[htb]
\centering
\includegraphics[width=0.9\columnwidth, angle=0]{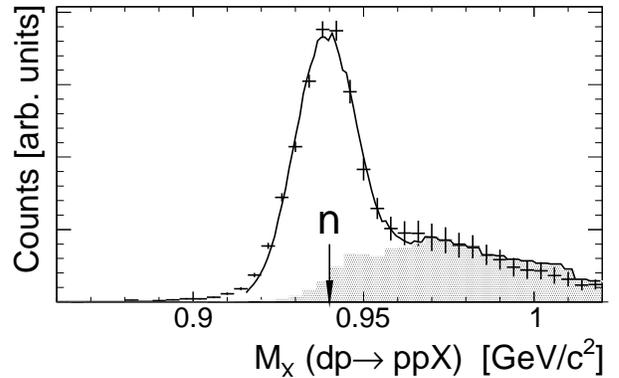}
\caption{Distribution in missing masses for the $dp\to ppX$ reaction for
726~MeV deuterons incident on a storage cell filled with polarised hydrogen.
Events that originated from the walls of the cell gave rise to the tail at
high $M_X$. The shape of this background was simulated by filling the cell
with nitrogen gas and this led to the shaded area. The solid line represents
the fit of a Gaussian plus the numerical values of the scaled background
histogram.}
\label{fig:Mx}
\end{figure}

The missing-mass $M_X$ distribution of the identified $dp\to ppX$ reaction
shows a striking peak around the mass of the missing neutron, as illustrated
in Fig.~\ref{fig:Mx}. The long tail to higher missing masses arises from
events originating from the walls of the storage cell. The shape of this
background was simulated by filling the cell with nitrogen gas. This gave
rise to the shaded area in the figure which, after normalising the
distribution at high $M_X$, could be reliably subtracted bin by bin. The
resulting $dp\to ppn$ events were then placed in 20~MeV/$c$ bins in the
momentum transfer $q$ between the deuteron and diproton and 2~MeV bins in the
diproton excitation energy $E_{pp}$.

For a vector polarised deuteron beam incident on a polarised hydrogen target,
where both polarisations are in the $y$ direction, the ratio of the numbers
of polarised $N(q,\phi)$ to unpolarised $N^{0}(q)$ events has the
form~\cite{OHL1972}:
\begin {eqnarray}
\nonumber \frac{N(q,\phi)}{N^{0}(q)} = 1 + p_p A_{y}^{p}(q)\cos\phi
+ \fmn{3}{2}p_d A_{y}^{d}(q)\cos\phi + \\
+ \fmn{3}{4}p_d
p_p[(1+\cos2\phi)\,C_{y,y}(q)+(1-\cos2\phi)\,C_{x,x}(q)],\phantom{1}
\label{eq:dc_general}
\end {eqnarray}
where the azimuthal angle $\phi$ is measured from the $x$-axis.\\

\begin{figure}[htb]
\centering
\includegraphics[width=0.9\columnwidth, angle=0]{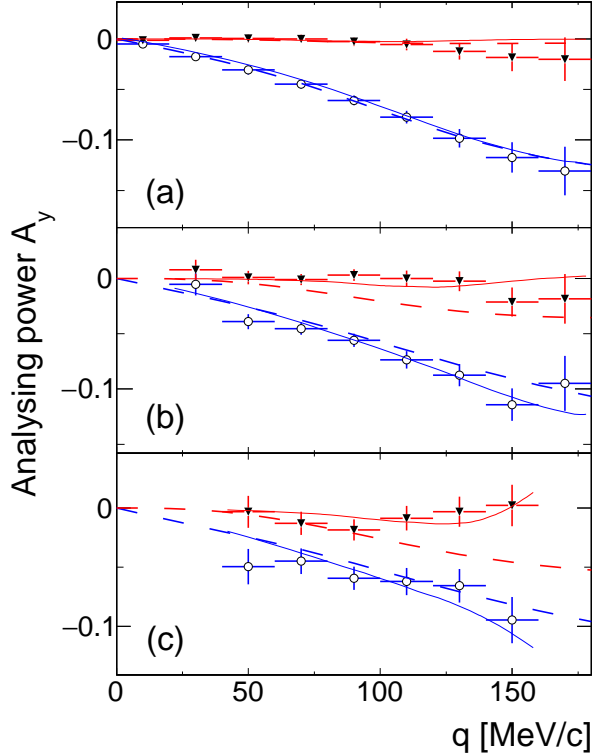}
\caption{Values of the deuteron (vector) analysing power $A_y^d$ (black
inverted triangles) and proton analysing power $A_y^p$ (blue open circles)
for the $dp\to\{pp\}n$ reaction at 726~MeV. Data were placed in bins of (a)
$E_{pp}<2$~MeV, (b) $4 < E_{pp}< 6$~MeV, and (c) $8 < E_{pp}< 10$~MeV; only
statistical errors are shown. The curves are impulse approximation
estimates~\cite{CAR1991} folded with experimental acceptance. These used the
current SAID neutron-proton partial wave solution~\cite{ARN2000} as input.
The dashed lines neglect the predicted dependence on the angle between the
relative momentum in the diproton and the momentum transfer. The model
predictions for this are included in the solid lines.} \label{fig:Ay}
\end{figure}

\begin{figure}[htb]
\centering
\includegraphics[width=0.9\columnwidth, angle=0]{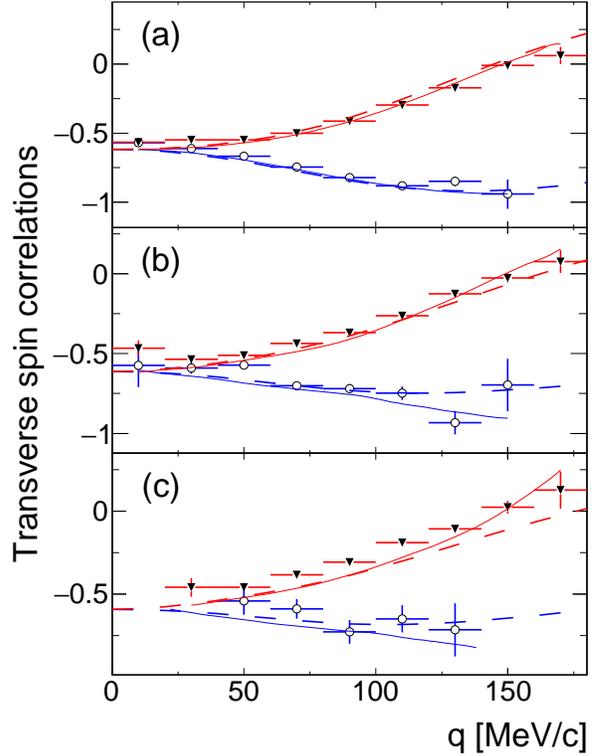}
\caption{Values of the spin correlations $C_{y,y}$ (black inverted triangles)
and $C_{x,x}$ (blue open circles) for the $dp\to\{pp\}_{\!s}n$ reaction at
726~MeV. The conventions are identical to those of Fig.~\ref{fig:Ay}.}
\label{fig:CxxCyy}
\end{figure}

From studying the $\phi$ dependence of the count rates for the four
combinations of beam and target polarisations it is possible to extract
separately the values of the proton and deuteron vector analysing powers as
well as the two spin correlations. The results for these observables are
shown in Figs.~\ref{fig:Ay} and \ref{fig:CxxCyy}. The deuteron vector
analysing power $A_y^d$ of Fig.~\ref{fig:Ay} remains very small over our
whole $q$ range, only (possibly) exceeding 1\% in magnitude for $q\gtrsim
120$~MeV/$c$. The proton analysing power $A_y^p$, though small, is much
larger than $A_y^d$. These two features are very similar to the results found
at 600~MeV per nucleon~\cite{MCH2013}, though the statistical precision of
the current data is much higher.

Due to the ANKE exit window being much wider in the horizontal direction than
in the vertical, for the larger values of $q$ the data are more populated
near $\sin\phi=0$. It follows from Eq.(\ref{eq:dc_general}) that, in this
limit, the spin correlation $C_{y,y}$ is better measured than $C_{x,x}$, and
this is seen in Fig.~\ref{fig:CxxCyy}. However, in order to assess the
significance of these results we must turn to a reaction model.

In impulse approximation the amplitude for the $dp\to\{pp\}n$ charge-exchange
reaction is proportional to the $np\to pn$ charge-exchange amplitude times a
form factor that represents the overlap of the initial deuteron wave function
with that of the outgoing diproton~\cite{BUG1987,CAR1991}. The elementary
$np\to pn$ amplitude may be written in terms of five scalar amplitudes in the
$np$ c.m.\ frame as:
\begin{eqnarray}
\nonumber f_{np}&=&\alpha(q) +i\gamma(q)
(\vec{\sigma}_{1}+\vec{\sigma}_{2})\cdot\vec{n} +\beta(q)
(\vec{\sigma}_{1} \cdot {\bf n})(\vec{\sigma}_{2}\cdot\vec{n})\\
&&+\delta(q)(\vec{\sigma}_{1}\cdot\vec{m})(\vec{\sigma}_{2}\cdot\vec{m})
+\varepsilon(q)(\vec{\sigma}_{1}\cdot\vec{l})(\vec{\sigma}_{2}\cdot\vec{l}),
 \label{fpn}
\end{eqnarray}
where $q=\sqrt{-t}$ is the three-momentum transfer between the initial
neutron and final proton and the Pauli matrices $\vec{\sigma}$ are sandwiched
between neutron and proton spinors. Of the unit basis vectors, $\vec{l}$ lies
along the mean of the initial proton and final neutron momenta, $\vec{n}$
lies along $\vec{q}$, and $\vec{m}=\vec{n}\times\vec{l}$. It should be noted
that the amplitudes of Eq.~(\ref{fpn}) are actually linear combinations of
the standard elastic ones, defined for example in Ref.~\cite{BYS1978}. This
important distinction arises because the spin dependence that is made
explicit here is that corresponding to charge
exchange~\cite{BUG1987,LEH2010}.

\begin{figure}[htb]
\centering
\includegraphics[width=0.9\columnwidth, angle=0]{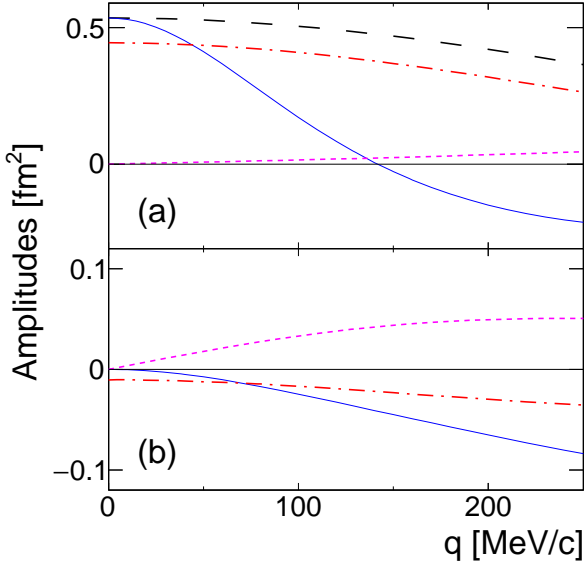}
\caption{Values of the (a) real and (b) imaginary parts of the $np$
amplitudes at 363~MeV predicted on the basis of the current SAID PWA
solution~\cite{ARN2000}. The amplitudes of Eq.~(\ref{fpn}), normalised to
$d\sigma/dq^2$, have been rotated such that $\beta(q)$ is real. The curves
are for $\beta$ (black), $\delta$ (blue), $\epsilon$ (red), and $\gamma$
(magenta). It should be noted that the scales in panels (a) and (b) are very
different.} \label{amps}
\end{figure}

Values of the amplitudes of Eq.~(\ref{fpn}) at 363~MeV can be extracted from
the current partial wave solution of the SAID group~\cite{ARN2000} and the
ones that are relevant to the current work are shown in Fig.~\ref{amps} as
functions of the momentum transfer $q$. Since only relative phases are
significant in the discussion, these amplitudes have been rotated in the
complex plane to make $\beta$ real for all $q$. Apart from the obvious
features that $\beta=\delta$ and $\gamma=0$ at $q=0$, the most notable
behaviour is the zero in the $\delta$ amplitude at $q\approx
(140-10i)$~MeV/$c$. This amplitude is strongly influenced by one pion
exchange and, in the simplest distorted model, this has a zero when
$q=m_{\pi}c$, where $m_{\pi}$ is the mass of the charged pion.

Although the resulting $dp\to\{pp\}n$ amplitudes will be evaluated taking
into account higher partial waves in the $pp$ system, using an update of the
program of Ref.~\cite{CAR1991}, it is useful for a qualitative discussion to
consider the results that follow if one retains only the $pp$ \Szero\
configuration that dominates at low $E_{pp}$. In this case the impulse
approximation model predicts that:
\begin{eqnarray}
\nonumber
A_y^d&=&0\:,\\
\nonumber
A_y^p&=&-2\textit{Im}(\beta^*\gamma)/(|\beta|^{2}+|\gamma|^{2}+|\varepsilon |^{2}+|\delta|^{2}),\\
\nonumber
C_{y,y}&=&-2\textit{Re}(\varepsilon^*\delta)/(|\beta|^{2}+|\gamma|^{2}+|\varepsilon |^{2}+|\delta|^{2}),\\
C_{x,x}&=&-2\textit{Re}(\varepsilon^*\beta)/(|\beta|^{2}+|\gamma|^{2}+|\varepsilon |^{2}+|\delta|^{2}).
\label{impulse}
\end{eqnarray}
In addition to taking the higher $pp$ partial waves into account, away from
$q=0$ these formulae have to be modified to include the effects of the
deuteron $D$-wave and the Wigner rotation that arises from a change in
reference frame~\cite{BUG1987}.

In the \Szero\ limit of the impulse approximation the spin correlations are
linked at $q=0$ to a combination of neutron-proton spin-correlation and
spin-transfer parameters, as defined in Ref.~\cite{BYS1978}, through
\begin{equation}
\label{prove}
C_{x,x}(0) = C_{y,y}(0) = \frac{2[A_{00nn}(\pi)-D_{n0n0}(\pi)]}
{3-K_{0ll0}(\pi)-2K_{0nn0}(\pi)}\cdot
\end{equation}

The curves shown in Figs.~\ref{fig:Ay}a and \ref{fig:CxxCyy}a represent the
full impulse approximation calculations for $E_{pp}< 2$~MeV~\cite{CAR1991}.
Though in this case the two protons should be dominantly in the \Szero\
configuration, there might still be some small deviation from the $A_y^d=0$
prediction of Eq.~(\ref{impulse}). However, both the data and the predictions
are at the 1\% level and it is hard to draw firm conclusions in view of the
systematic uncertainties. The situation is much clearer for the proton
analysing power of Fig.~\ref{fig:Ay}a, where the current SAID solution
provides a quantitative description of the experimental data. Since the real
and imaginary parts of $\gamma$ in Fig.~\ref{amps} are of comparable size,
this suggests that the phase between $\beta$ and $\gamma$ is well reproduced
in the partial wave solution~\cite{ARN2000}.

The general shapes of the $C_{x,x}$ and $C_{y,y}$ predictions in
Fig.~\ref{fig:CxxCyy}a are very much in line with the measured data though
there are some quantitative differences. $C_{y,y}$ changes sign, as one would
expect from the simple one-pion-exchange contribution to the $\delta$
amplitude, though this happens at a few MeV/$c$ higher than the prediction.
While being independent of the uncertainties in the beam and target
polarisations, this crossing is a very sensitive test of the
$\delta/\epsilon$ interference because of the presence of the small imaginary
part in the zero of the $\delta$ amplitude at $q\approx(140-10i)$~MeV/$c$.

The prediction of $C_{x,x}(0)\approx-0.62$ at $q=0$ is to be compared to the
value of $\approx -0.57$ extracted from the experimental data for $E_{pp}<
2$~MeV. The significance of the discrepancy here must be judged against the
systematic uncertainties in the experiment and the modeling. Systematic
effects in the data could arise, for example, from the choice in fitting
limits combined with some imperfection in the background description, but
these can be estimated conservatively to be below 0.01 for the analysing
powers and 0.03 for the spin correlations. To these must be added the 4\%
associated with the product of the beam and target polarisations. The
uncertainties arising from the beam or target polarisations are small
compared to the statistical errors in $A_y^d$ and $A_y^p$.

The SAID single-energy solution yields error bars on the $np$ spin-transfer
and correlation parameters needed to evaluate Eq.~(\ref{prove}). The
resulting uncertainty in $C_{x,x}(0)$ of $\pm 0.024$~\cite{WOR2014} is very
much a lower limit because it does not include any uncertainties in the model
assumed in the SAID analysis or in the data selection~\cite{ARN2000}. It must
also be stressed that few of the four $np$ observables appearing in
Eq.~(\ref{prove}) have been directly measured near $180^{\circ}$.
Furthermore, the SAID predictions do show very strong angular dependence,
which makes any extrapolation in angle less reliable. On the other hand, the
uncertainty of 4\% in the product of the beam and target polarisations would
correspond to a systematic error of $\pm 0.025$ in the determination of
$C_{x,x}(0)$.

The $\pol{d}\pol{p}\to \{pp\}_{\!s}n$ data were here described using a
plane-wave impulse approximation~\cite{BUG1987,CAR1991} and the largest
correction to this picture is likely to come from double scattering inside
the deuteron. Evaluating the \Szero\ contribution in an eikonal approach, it
has been shown that the spin dependence of the deuteron tensor analysing
powers is little changed by this correction for $q\lesssim
140$~MeV/$c$~\cite{WIL2014}. An estimation of the double
scattering~\cite{BUG1987,WIL2014} at $q=0$ indicates that this modifies the
prediction for $C_{x,x}(0)$ by only $0.003$. This is an order of magnitude
less than the quoted uncertainties and so can be safely neglected.

For larger values of $E_{pp}$, $P$ and higher waves become significant so
that the angular distribution in the diproton is no longer isotropic. If
$\vec{k}$ is the relative momentum in the diproton, then $E_{pp}=k^2/m$,
where $m$ is the proton mass. At large values of $k$ and $q$, where the
effects of the Pauli exclusion principle and final state interactions are
small, there will be a quasi-free peak at $\vec{k}=\pm\vec{q}/2$. Quite
generally therefore, away from the small $E_{pp}$ region there will be a
significant dependence on the angle $\theta_{kq}$ between $\vec{k}$ and
$\vec{q}$~\cite{BUG1987}. It is important to note that this effect is already
included in the computer program of Ref.~\cite{CAR1991}.

Panels b and c of Figs.~\ref{fig:Ay} and \ref{fig:CxxCyy} show the
experimental results obtained in the bins $4 < E_{pp} < 6$~MeV and $8 <
E_{pp} < 10$~MeV, respectively. The broken lines indicate the plane wave
impulse approximation prediction where one ignores the angular distribution
in $\theta_{kq}$. When the predictions~\cite{CAR1991} of this angular
dependence are included one obtains the solid lines in the figures. Though
most of the changes are small compared to the uncertainties in the $np$ input
data, these generally go in the right direction, especially for the deuteron
vector analysing power. The same is also true for the results in two $E_{pp}$
bins that are not shown here.

Spin correlations and analysing powers have been measured in deuteron charge
exchange on hydrogen, $\pol{d}\pol{p}\to \{pp\}n$, at a beam energy of
726~MeV. The high statistics of this experiment allowed tight cuts to be
placed on the $pp$ excitation energy $E_{pp}$. The agreement of these data
with the impulse approximation model at very low $E_{pp}$, where the \Szero\
state will be dominant, shows that the $np$ amplitudes obtained from partial
wave analysis~\cite{ARN2000} must be broadly correct. There are slight
discrepancies near the forward direction but these are of such a size that
they could originate from the partial wave solution or from uncertainties in
the experimental data presented here.

At larger values of $E_{pp}$ there are small effects associated with $P$ and
higher partial waves in the $pp$ system that lead to some non-isotropy in the
diproton angular distribution. These were studied by putting the data in five
2~MeV bins in $E_{pp}$. Though the acceptance in ANKE is less complete at
large vales of $E_{pp}$, the plane wave impulse approximation model describes
all these effects.

\newpage

We are grateful to other members of the ANKE Collaboration for their help
with this experiment and to the COSY crew for providing such good working
conditions, especially in respect of the polarised beam. The values of the
SAID neutron-proton amplitudes were kindly furnished by I.I.~Strakovsky. This
work has been partially supported by the Forschungszentrum J\"ulich COSY-FFE
\#73 and \#80, and the Georgian National Science Foundation.

%
%

%

\begin{thebibliography}{99}
%
\bibitem{MCG1960} M.H.~McGregor, M.J.~Moravcsik, and H.P.~Stapp, Ann.\ Rev.\
    Nucl.\ Sci.\ 10 (1960) 291.
%
\bibitem{BYS1978} J.~Bystricky, F.~Lehar, P.~Winternitz,
    J.\ Phys.\ (Paris) 39 (1978) 1.
%
\bibitem{BUG1987} D.V.~Bugg, C.~Wilkin, Nucl.\ Phys.\ A
    467 (1987) 575.
%
\bibitem{MCH2013} D.~Mchedlishvili et al., Eur.\ Phys.\ J.\ A 49 (2013) 49.
%
\bibitem{ARN2000} R.A.~Arndt, I.I.~Strakovsky, R.L.~Workman,
    Phys.\ Rev.\ C {62} (2000) 034005;
 R.A.~Arndt, W.J.~Briscoe, I.I.~Strakovsky,\\ R.L.~Workman, Phys.\ Rev.\ C {76} (2007)
 025209;\\ \verb=http://gwdac.phys.gwu.edu=.
%
\bibitem{CAR1991} J.~Carbonell, M.B.~Barbaro, C.~Wilkin,
    Nucl.\ Phys.\ A 529 (1991) 653.
%
\bibitem{BAR1997} S.~Barsov et al., Nucl.\ Instrum.\ Meth.\  A
    462 (1997) 364.
%
\bibitem{DYM2013} S.~Dymov et al., Phys.\ Rev.\ C 88 (2013) 014001.
%
\bibitem{DYM2004} S.~Dymov et al., Part.\ Nucl.\ Lett.\  {2 (119)} (2004) 40.
%
\bibitem{OHL1972} G.G.~Ohlsen, Rep.\ Prog.\ Phys.\ 35 (1972) 717.
%
\bibitem{LEH2010} F.~Lehar, C.~Wilkin, Phys.\ Part.\
    Nuclei Lett.\ 7 (2010) 235.
%
\bibitem{WOR2014} R.~Workman, \emph{private communication} (2014).
%
\bibitem{WIL2014} C.~Wilkin in \emph{NN and 3N Interactions}, Ed.
    L.~Blokhintsev and I.I.~Strakovsky, (Nova Science, N.Y., 2014).
%
\end{thebibliography}
\end{document}